\documentclass[twocolumn,showpacs,superscriptaddress,10pt]{revtex4-1}
\usepackage{graphicx,color,amssymb,amsthm,hyperref}
\usepackage{dcolumn}
\usepackage{bm}
\usepackage{bbm}
\usepackage{amsmath}
\usepackage{soul}

\usepackage{multirow,array,rotating}
\newcolumntype{M}{>{$\vcenter\bgroup\hbox\bgroup}c<{\egroup\egroup$}}
\newcommand{\ve}{\boldsymbol}

\newcommand{\dd}{\mathop{}\!\mathrm{d}}

\def\>{\rangle}
\def\<{\langle}

\def\I{ \mathbbm{1} }

\DeclareMathOperator{\tr}{tr}
\DeclareMathOperator{\diag}{diag}

\def\diag{ \mbox{diag} }

\newtheorem{thm}{Theorem}

\definecolor{nblue}{rgb}{0.3,0.3,1.0}
\definecolor{ngreen}{rgb}{0.2,0.7,0.2}
\definecolor{nred}{rgb}{0.9,0.1,0}
\definecolor{norange}{rgb}{0.8,0.5,0}

\newcommand{\sch}{Schr\"odinger}

\begin{document}

\title{Einstein-Podolsky-Rosen steering and the steering ellipsoid}

\author{Sania Jevtic}
\affiliation{Mathematical Sciences, John Crank 501, Brunel University, Uxbridge UB8 3PH, United Kingdom}
\author{Michael J. W. Hall}
\affiliation{Centre for Quantum Computation and Communication Technology (Australian Research Council), Centre for Quantum Dynamics, Griffith University, Brisbane, QLD 4111, Australia}
\author{ Malcolm R. Anderson}
\affiliation{Mathematical and Computing Sciences, Universiti Brunei Darussalam,
Gadong BE 1410, Negara Brunei Darussalam}
\author{Marcin Zwierz}
\affiliation{Faculty of Physics, University of Warsaw, ul. Pasteura 5, PL-02-093 Warszawa, Poland}
\affiliation{Centre for Quantum Computation and Communication Technology (Australian Research Council), Centre for Quantum Dynamics, Griffith University, Brisbane, QLD 4111, Australia}
\author{Howard M. Wiseman}
\affiliation{Centre for Quantum Computation and Communication Technology (Australian Research Council), Centre for Quantum Dynamics, Griffith University, Brisbane, QLD 4111, Australia}

\begin{abstract}
The question of which two-qubit states are steerable (i.e. permit a demonstration of EPR-steering) remains open.
Here, a strong necessary condition is obtained for the steerability of two-qubit states having maximally-mixed reduced states, via the construction of local hidden state models.
It is conjectured that this condition is in fact sufficient.  Two provably sufficient conditions are also obtained, via asymmetric EPR-steering  inequalities.
Our work uses ideas from the quantum steering ellipsoid formalism, and explicitly evaluates the integral of $\ve n/(\ve n^\intercal A\ve n)^2$ over arbitrary  unit  hemispheres for any positive matrix $A$.
\end{abstract}


\maketitle

\section{Introduction}

Quantum systems can be correlated in ways that supersede classical descriptions. However,
there are degrees of  non-classicality  for quantum correlations.  For simplicity, we consider
only bipartite correlations, with the two, spatially separated, parties being named Alice and Bob as usual.

At the weaker end of the spectrum are quantum systems
whose states cannot be expressed as a mixture of product-states of  the  constituents.
These are called non-separable or entangled states.
The product-states appearing in such a mixture comprise a local hidden state (LHS) model for any
measurements undertaken by Alice and Bob.

At the strongest end of the spectrum are quantum systems whose measurement correlations can violate a Bell inequality~\cite{bell64,chsh},
hence demonstrating (modulo loopholes~\cite{larsson14}) the violation of local causality~\cite{bell76}.
This phenomenon---commonly known as Bell-nonlocality~\cite{wiseman14}---is the only way for two spatially separated parties to verify the existence of entanglement if either of them, or their detectors, cannot be trusted~\cite{terhal00}. We say that a bipartite state is Bell-local if and only if there is a local hidden variable (LHV) model for any measurements Alice and Bob perform. Here the `variables'
are not restricted to be quantum states, hence the distinction between non-separability and Bell-nonlocality.

In between these types of non-classical correlations lies EPR-steering. The name is inspired by the seminal paper
of Einstein, Podolsky, and Rosen (EPR)~\cite{einstein35}, and the follow-up by \sch~\cite{schrsteer35}, which coined the term ``steering''
for the phenomenon EPR had noticed. Although introduced eighty years ago, as this Special Issue celebrates, the notion of EPR-steering
was only formalized eight years ago, by one of us and co-workers~\cite{wiseman07,jones07}.
This formalization was that EPR-steering is the only way to verify the existence of entanglement if
one of the parties --- conventionally Alice~\cite{wiseman07,jones07,cavalcanti09} ---
or her detectors, cannot be trusted.
We say that a bipartite state is EPR-steerable if and only if it allows a demonstration of EPR-steering.
A state is not EPR-steerable if and only if there exists a hybrid LHV--LHS model explaining
the Alice--Bob correlations. Since in this paper we are concerned with steering,
when we refer to a LHS model we mean a LHS model for Bob only; it is implicit
that Alice can have a completely general LHV model.

The above three notions of non-locality for quantum states coincide for pure states:
 any non-product pure state is non-separable, EPS-steerable, and Bell-nonlocal.
However for mixed states, the interplay of quantum and classical correlations produces a far richer structure.
 For mixed states the logical hierarchy of the three concepts leads to a hierarchy for the bipartite states:
the set of separable states is a strict subset of the set of  non-EPR-steerable states, which is a strict subset of the set of Bell-local states~\cite{wiseman07,jones07}.

Although the EPR-steerable set has been completely determined for certain classes of highly symmetric states (at least for the case
where Alice and Bob perform projective measurements)~\cite{wiseman07,jones07}, until now very little was known
about what types of states are steerable even for the simplest case of two qubits. In this simplest case,
the phenomenon of steering in a more general sense --- i.e. within what set can Alice steer Bob's state by measurements on her system ---
has been studied extensively using the so-called steering ellipsoid formalism \cite{verstraete2002,Shi,QSE}. However, no relation between the steering ellipsoid and EPR-steerability has been determined.

In this manuscript, we investigate EPR-steerability of the class of two-qubit states whose reduced states are maximally mixed,
the so-called T-states~\cite{Tstates}. We use the steering ellipsoid formalism to develop a deterministic LHS model for  projective measurements on these states and we conjecture that this model is optimal. Furthermore we obtain two sufficient conditions for T-states to be EPR-steerable, via suitable EPR-steering inequalities~\cite{cavalcanti09, jones11} (including a new asymmetric steering inequality for the spin covariance matrix).
These sufficient conditions touch the necessary condition in some regions of the space of T-states, and everywhere else the gap between
them is quite small.

The paper is organised as follows. In section 2 we discuss in detail the three notions of non-locality, namely Bell-nonlocality, EPR-steerability and non-separability. Section 3 introduces the quantum steering ellipsoid formalism for a two-qubit state, and in section 4 we use the steering ellipsoid to develop a deterministic LHS model for projective measurements on T-states. In section 5, two asymmetric steering inequalities for arbitrary two-qubit states are derived. Finally in section 6 we conclude and discuss further work.

\section{\label{sec_LHSmodel}EPR-steering and local hidden state models}

Two separated observers, Alice and Bob, can use a shared  quantum state to generate statistical correlations between local measurement outcomes.  Each observer carries out a local measurement, labelled by $A$ and $B$ respectively, to obtain corresponding outcomes labelled by $a$ and $b$. The measurement correlations are described by some set of joint probability distributions, $\{p(a,b|A,B)\}$, with $A$ and $B$ ranging over the available measurements. The type of state shared by Alice and Bob may be classified via the properties of these joint distributions,
for all possible measurement settings $A$ and $B$.

The correlations of a {\it Bell-local} state have a local hidden variable (LHV) model \cite{bell64, chsh},
\begin{equation} \label{lhv}
p(a,b|A,B) = \sum_\lambda P(\lambda) \,p(a|A,\lambda)\,p(b|B,\lambda) ,
\end{equation}
for some `hidden' random variable $\lambda$ with probability distribution $P(\lambda)$.  Hence, the measured correlations may be understood as arising from ignorance of the value of $\lambda$, where the latter locally determines the statistics of the outcomes $a$ and $b$ and is independent of the choice of $A$ and $B$. Conversely, a state is defined to be Bell{\it-nonlocal} if it has no LHV model.  Such states allow, for example, the secure generation of a cryptographic key between Alice and Bob without trust in their devices~\cite{qkd,qkd2}.

In this paper, we are concerned with whether the state is {\em steerable}; that is, whether it allows for correlations
that demonstrate EPR-steering. As discussed in the introduction, EPR-steering by Alice is demonstrated if it is {\em not} the case
that the correlations can be described by a hybrid LHV--LHS model, wherein,
\begin{equation} \label{lhs}
p(a,b|A,B) =  \sum_\lambda P(\lambda) \,p(a|A,\lambda)\,p_Q(b|B,\lambda),
\end{equation}
where the local distributions $p_Q(b|B,\lambda)$ correspond to measurements on local quantum states $\rho_B(\lambda)$, i.e.,
\[ p_Q(b|B,\lambda) = {\rm tr}[\rho_B(\lambda) F^B_b ].  \]
Here $\{ F^B_b\}$ denotes the positive operator valued measure (POVM) corresponding to measurement $B$. The state is said to be {\it steerable} by Alice if there is {\it no} such model. The roles of Alice and Bob may also be reversed in the above, to define steerability by Bob.

Comparing Eqs.~(\ref{lhv}) and (\ref{lhs}), it is seen that all nonsteerable states are Bell-local.  Hence, all Bell-nonlocal states are steerable, by both Alice and Bob. In fact, the class of steerable states is strictly larger \cite{wiseman07}.  Moreover, while not as powerful as Bell-nonlocality in general, steerability is more robust to detection inefficiencies \cite{ineff}, and also enables the use of untrusted devices in
quantum key distribution, albeit only on one side \cite{steeringqkd}.  By a similar argument, a separable quantum state shared by Alice and Bob, $\rho=\sum_\lambda p(\lambda) \rho_A(\lambda) \otimes \rho_B(\lambda)$, is both Bell-local and nonsteerable. Moreover, the set of separable states is strictly smaller than the set of nonsteerable  states~\cite{wiseman07}.

It is important that EPR-steerability of a quantum state not be confused with merely the dependence of the reduced state of one observer on the choice of measurement made by another, which can occur even for separable states.
The term `steering' has been used with reference to this phenomenon, in particular for the concept of `steering ellipsoid', which we will use in our analysis. EPR-steering, as defined above, is a special case of this phenomenon, and is only possible for a subset of nonseparable states.

We are interested in the EPR-steerability of states for all possible \textit{projective} measurements. If Alice is doing the steering, then it is sufficient for Bob's measurements to comprise some tomographically complete set of projectors.  It is straightforward to show in this case that the condition for Bob to have an LHS model, Eq.~(\ref{lhs}), reduces to the existence of a representation of the form
\begin{subequations}
\begin{align}
\label{reduced}
\hspace{-0.3cm}
p_E\, \rho^E_B&:= {\rm tr}_A[\rho \,E\otimes \I] = \sum_\lambda P(\lambda)\, p(1|E,\lambda) \,\rho_B(\lambda) ,\\
\label{reduced_p}
p_E &= \tr [ \rho E\otimes I] = \sum_\lambda P(\lambda) p(1|E,\lambda).
\end{align}
\end{subequations}
Here $E$ is any  projector that can be measured by Alice; $p_E$ is the probability of result `$E=1$' and $p(1|E,\lambda)$ is the corresponding probability given $\lambda$; $\rho^E_B$ is the reduced state of Bob's component corresponding to this result; and ${\rm tr}_A[\cdot]$ denotes the partial trace over Alice's component. Note that this form, and hence EPR-steerability by Alice, is invariant under local unitary transformations on Bob's components.

Determining EPR-steerability in this case, where  Alice is permitted to measure any Hermitian observable, is surprisingly difficult, with the answer only known for certain special cases such as Werner states \cite{wiseman07}.  However, in this paper we give a strong necessary condition for the EPR-steerability of a large class of two-qubit states, which we conjecture is also sufficient.  This condition is obtained via the construction of a suitable LHS model, which is in turn motivated by properties of the `quantum steering ellipsoid'~\cite{verstraete2002, QSE}. Properties of this ellipsoid are therefore reviewed in the following section.

\section{\label{sec_QSE}The quantum steering ellipsoid}

An arbitrary two-qubit state may be written in the standard form
\begin{align*}
\rho = \frac{1}{4}\left(\I\otimes\I+\ve{a}\cdot \ve{\sigma}\otimes \I + \I\otimes \ve{b}\cdot\ve{\sigma} + \sum_{j,k} T_{jk} \,\sigma_j\otimes\sigma_k\right).
\end{align*}
Here $(\sigma_1,\sigma_2,\sigma_3)\equiv\ve{\sigma}$ denote the Pauli spin operators, and
$$ a_j=\tr [\rho \,\sigma_j\otimes\I],~ b_j=\tr [\rho\,  \I\otimes\sigma_j],~T_{jk}=\tr [\rho \,\sigma_j\otimes\sigma_k] . $$
Thus, $\ve{a}$ and $\ve{b}$ are the Bloch vectors for Alice and Bob's qubits, and $T$ is the spin correlation matrix.

If Alice makes a  projective measurement on her qubit, and obtains an outcome corresponding to  projector $E$,  Bob's reduced state follows from  Eq.~(\ref{reduced}) as
\[  \rho^E_B =\frac{\tr_A [\rho\, E \otimes \I] }{\tr[\rho \,E \otimes \I] }. \]
We will also refer to $\rho^E_B$ as Bob's `steered state'.

To determine Bob's possible steered states, note that the projector  $E$ may be expanded in the Pauli basis as $E = \frac12  \left( \I + \ve{e}\cdot\ve{\sigma}\right)$, with $|\ve e|= 1$. This yields the corresponding steered state
$ \rho^E_B = \frac 12\left(\I + \ve{b}(\ve e)\cdot\ve{\sigma} \right)$,
with associated Bloch vector
\begin{equation} \label{qsteerB}
\ve{b}(\ve e) = \frac{1}{2p_{\ve e}} (\ve b + T^\intercal\ve{e}),
\end{equation}
where $p_{\ve e}$  is the associated probability of result $`E=1$',
\begin{equation} \label{pe}
p_{\ve e}  := \tr[\rho (E \otimes \I)] = \frac 12 (1+\ve a\cdot\ve e),
\end{equation}
called $p_E$ previously. In what follows we will refer to the vector $\ve e$ rather than its corresponding operator $E$.

The surface of the steering ellipsoid is  defined to be  the set of steered Bloch vectors, $\{ \ve{b}(\ve e): |\ve e|=1\}$, and in Ref.~\cite{QSE} it is shown that interior points can be obtained from positive operator-valued measurements (POVMs). The ellipsoid has centre
\begin{equation}
\label{QSE_centre}
\ve c = \frac{\ve b - T^\intercal\ve a}{1-a^2} ,
\end{equation}
and the semiaxes  $s_1,s_2, s_3$ are the roots of the eigenvalues of the matrix
\begin{equation}
\label{QSE_mat}
Q = \frac{1}{1-a^2}\left(T^\intercal - \ve b \ve a^\intercal \right)\left(\I + \frac{\ve a \ve a^\intercal}{1-a^2} \right)\left(T - \ve a \ve b ^\intercal \right) .
\end{equation}
The eigenvectors of $Q$ give the orientation of the ellipsoid around its centre \cite{QSE}. Thus, the general equation of the steering ellipsoid surface is $\ve x^\intercal Q^{-1}\ve x=1$ with $\ve x \in \mathbb{R}^3$ being the displacement vector from the centre $\ve{c}$.

Entangled states  typically have large steering ellipsoids---the largest possible being the Bloch ball, which is generated by every pure entangled state \cite{QSE}.  In contrast, the volume of the steering ellipsoid is strictly bounded  for separable states. Indeed, a two-qubit state is separable if and only if its steering ellipsoid is contained within a tetrahedron contained within the Bloch sphere \cite{QSE}. Thus, the separability of two-qubit states has a beautiful geometric characterisation in terms of the quantum steering ellipsoid.

No similar characterisation has been found for EPR-steerability, to date. However, for non-separable states, knowledge of the steering ellipsoid matrix $Q$, its centre $\ve c$, and Bob's Bloch vector $\ve b$ uniquely determines the shared state $\rho$ up to a local unitary transformation on Alice's system \cite{QSE}, \cite{Note1} and so is sufficient, in principle, to determine the EPR-steerability of $\rho$. In this paper we find a direct connection between EPR-steerability and the quantum steering ellipsoid, for the case that the Bloch vectors $\ve a$ and $\ve b$ vanish.

\section{\label{sec_Tstate}Necessary condition for EPR-steerability of T-states}

\subsection{T-states}

Let $T=O_A \widetilde D O_B^\intercal$ be a singular value decomposition of the spin correlation matrix $T$,  for some diagonal matrix $\widetilde D\geq 0$ and orthogonal matrices $O_A,O_B \in \mathrm{O}(3)$.  Noting that any $O\in \mathrm{O}(3)$ is either a rotation or the product of a rotation with the parity matrix $-I$, it follows that $T$ can always be represented in the form $T=R_A D R_B^\intercal$, for proper rotations $R_A, R_B \in \mathrm{SO}(3)$, where the diagonal matrix $D$ may now have negative entries.

The rotations $R_A$ and $R_B$ may be implemented by local unitary operations on the shared state $\rho$, amounting to a local basis change.
Hence, all properties of a shared two-qubit state, including steerability properties in particular, can be formulated in a representation in which the spin correlation matrix has the diagonal form
$T \equiv D = {\rm diag}[t_1, t_2,t_3]$.
It follows that if the shared state $\rho$ has maximally-mixed reduced states with $\ve a=\ve b=\ve 0$, then it is completely described, up to local unitaries, by a diagonal $T$, i.e. one may consider
\begin{align}
\rho = \frac{1}{4} \left( \I\otimes \I + \sum_j t_j \sigma_j\otimes \sigma_j \right)
\end{align}
without loss of generality. Such states are called T-states~\cite{Tstates}. They are equivalent to mixtures of Bell states, and hence form a tetrahedron in the space parameterised by $(t_1,t_2,t_3)$ \cite{Tstates}. Entangled T-states necessarily have $t_1 t_2 t_3 < 0$,  and the set of separable T-states forms an octahedron within the tetrahedron \cite{Tstates}.

The T-state steering ellipsoid is centred at the origin, $\ve c = \ve 0$, and the ellipsoid matrix is simply $Q = T^\intercal T$, as follows from Eqs.~(\ref{QSE_centre}) and (\ref{QSE_mat}) with $\ve a= \ve b= \ve 0$. The semiaxes are $s_i = |t_i|$ for $i=1,2,3$, and are aligned with the $x,y,z$-axes of the Bloch sphere. Thus, the equation of the ellipsoid surface in spherical coordinates  $(r,\theta,\phi)$ is $r=1/f(\theta,\phi)$, with
\begin{align}
f(\theta,\phi)^2 := \frac{\sin^{2}\theta\cos ^{2}\phi}{s_1^2}  + \frac{\sin ^{2}\theta\sin ^{2}\phi}{s_2^2}  +\frac{\cos^{2}\theta}{s_3^2}.
\label{ell_eqn}
\end{align}
We find a remarkable connection between this equation and the EPR-steerability of T-states in the following subsection.

\subsection{\label{sec_Tstate_results}Deterministic LHS models for T-states}

Without loss of generality,
consider measurement by Alice of Hermitian observables on her qubit.  Such observables can be equivalently represented via projections, $E=\frac12 (\I+\ve e.\cdot\ve\sigma)$, with $|\ve e|=1$.  The probability of result `$E=1$' and the corresponding steered Bloch vector are given by Eqs.~(\ref{qsteerB}) and (\ref{pe}) with $\ve a=\ve b=\ve 0$, i.e.,
\[ p_{\ve e}=1/2,~~~~~~\ve b(\ve e) = T^T\ve e=T\ve e. \]
Hence, letting $\ve n(\lambda)$ denote the Bloch vector corresponding to $\rho_B(\lambda) $ in  Eq.~(\ref{reduced}), then from Eqs.~(\ref{reduced}) and \eqref{reduced_p}, it follows  there is an LHS model for Bob  if and only if there is a representation of the form
\[ \sum_\lambda P(\lambda)\,p(1|\ve e, \lambda) = \frac12, \!~~~\sum_\lambda P(\lambda)\,p(1|\ve e, \lambda)\,\ve n(\lambda) = \frac12 T\ve e,\]
for all unit vectors $\ve e$.  Noting further that $\ve n(\lambda)$ can always be represented as some mixture of unit vectors, corresponding to pure $\rho_B(\lambda)$, these conditions are equivalent to the existence of a representation of the form
\begin{align} \label{peint}
\int P(\ve n) \,p(1|\ve e,\ve n)\dd^2\ve n &= \frac12,\\
\int P(\ve n) \,p(1|\ve e, \ve n)\, \ve n\dd^2\ve n &= \frac12 T\ve e, \label{beint}
\end{align}
with integration over the Bloch sphere.  Thus, the unit Bloch vector $\ve n$ labels both the local hidden state and the hidden variable.

Given LHS models for Bob  for any two T-states, having spin correlation matrices $T_0$ and $T_1$, it is trivial to construct an LHS model for the T-state corresponding to $T_q=(1-q)T_0+qT_1$, for any $0\leq q\leq 1$, via the convexity property of nonsteerable states \cite{cavalcanti09}.
Our strategy is to find {\it deterministic} LHS models for some set of T-states, for which the result `$E=1$' is fully determined by knowledge of $\ve n$, i.e.,  $p(1|\ve e, \ve n)\in \{0,1\}$.  LHS models can then be constructed for all convex combinations of T-states in this set.

To find deterministic LHS models, we are guided by the fact that the steered Bloch vectors $\ve b(\ve e)=T\ve e$ are precisely those vectors that generate the surface of the quantum steering ellipsoid for the T-state \cite{QSE}. We make the ansatz that $P(\ve n)$ is proportional to some power of the function $f(\theta,\phi)$ in Eq.~(\ref{ell_eqn}) that defines this surface, i.e.,
\begin{equation} \label{pform}
P(\ve n) = N_T  \left[ f(\theta,\phi)\right]^m  \equiv N_T\,\left[\ve n^\intercal T^{-2}\ve n\right]^{m/2}
\end{equation}
for $\ve n=(\sin\theta\cos\phi,\sin\theta\sin\phi,\cos\theta)$, where $N_T$ is a normalisation constant. Further, denoting the region of the Bloch sphere, for which $p(1|\ve e,\ve n)=1$ by $\mathcal{R}[\ve e]$, the condition in Eq.~(\ref{peint}) becomes
$\int_{\mathcal{R}[\ve e]} P(\ve n) \dd^2\ve n = \frac12$.
We note this is automatically satisfied if $R(\ve e)$ is a hemisphere, as a consequence of the symmetry $P(\ve n)=P(-\ve n)$ for the above form of $P(\ve n)$.

Hence, under the assumptions that (i) $P(\ve n)$ is determined by the steering ellipsoid as per Eq.~(\ref{pform}), and (ii) $\mathcal{R}[\ve e]$ is a hemisphere for each unit vector $\ve e$, the only remaining constraint to be satisfied by a deterministic LHS model for a T-state is
Eq.~(\ref{beint}), i.e.,
\begin{equation} \label{con}
N_T \int_{\mathcal{R}[\ve e]}  \left[ \ve n^\intercal T^{-2} \ve n\right]^{m/2}\,\ve n \dd^2\ve n = \frac12 T\ve e,
\end{equation}
for some suitable mapping $\ve e\rightarrow \mathcal{R}[\ve e]$.

Extensive numerical testing, with different values of the exponent $m$, show that this constraint can be satisfied by the choices
\begin{equation} \label{re}
m=-4,~~~~~~~~\mathcal{R}[\ve e]= \{\ve n: \ve nT^{-1} \ve e\geq 0\},
\end{equation}
for a two-parameter family of T-states.  Assuming the numerical results are correct, it is not difficult to show, using infinitesimal rotations of $\ve e$ about the $z$-axis, that this family corresponds to those T-states that satisfy
\begin{equation} \label{surface}
2\pi N_T |\det T| = 1.
\end{equation}
Fortunately, we have been able to confirm these results analytically by explicitly evaluating the integral in Eq.~(\ref{con}) for $m=-4$ (see Appendix A). An explicit form for the corresponding normalisation constant $N_T$ is also given in Appendix A, and it is further shown  that the family of T-states satisfying Eq.~(\ref{surface}) is equivalently defined by the condition
\begin{equation} \label{surface2}
\int \sqrt{\ve n^\intercal T^2 \,\ve n} \dd^2\ve n =  2\pi.
\end{equation}
This may be interpreted geometrically in terms of the harmonic mean
radius of the `inverse' ellipsoid  $\ve x^\intercal\, T^2\ve x=1$ being equal to $2$.

\subsection{Necessary EPR-steerability condition}

Equation (\ref{surface}) defines a surface in the space of possible $T$ matrices, plotted in Fig.~1(a) as a function of the semiaxes $s_1, s_2$ and $s_3$.
As a consequence of the convexity of nonsteerable states (see above), all T-states corresponding to the region defined by this surface and the positive octant have local hidden state models for Bob.  Also shown is  the boundary of  the separable T-states ($s_1+s_2+s_3\leq 1$ \cite{Tstates}),  in red, which is clearly a strict subset of the nonsteerable T-states. The green plane corresponds to the sufficient condition $s_1+s_2+s_3 > \frac{3}{2}$ for EPR-steerable states, derived in Sec.~5 below.

It follows that a necessary condition for a T-state to be EPR-steerable by Alice is that it corresponds to a point above the sandwiched surface shown in Fig.~1(a).  Note that this condition is in fact symmetric between Alice and Bob, since their steering ellipsoids are the same for T-states. Because of the elegant relation between our LHS model and the steering ellipsoid, and other evidence given below, we conjecture that this condition is also {\it sufficient} for EPR-steerability.

\begin{figure}[t]
\begin{center}
\includegraphics[width=3.5in]{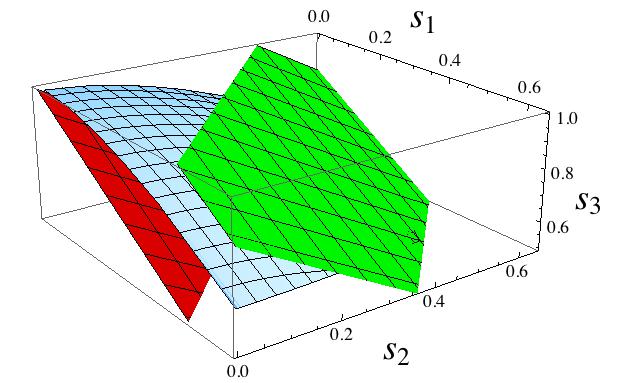}
\includegraphics[width=2.7in]{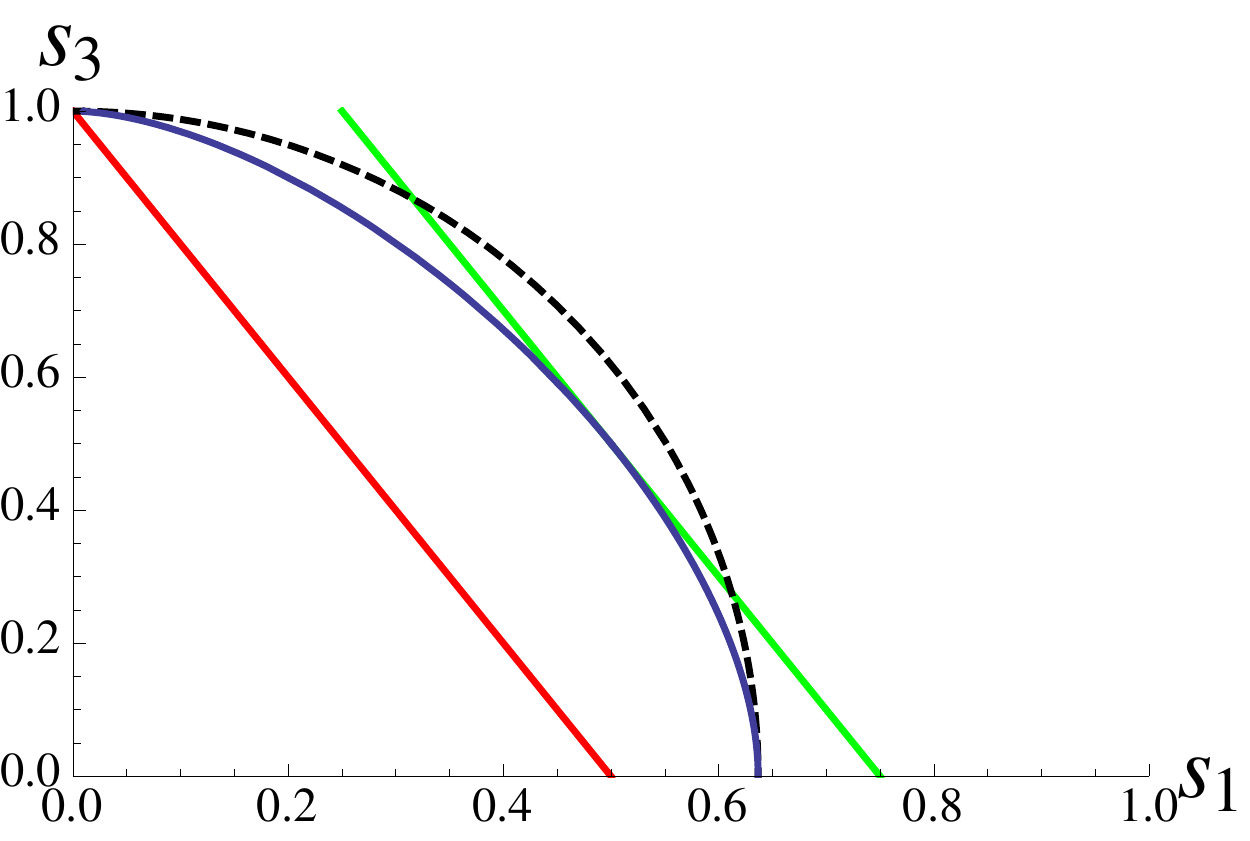}
\caption{Correlation  bounds for  T-states, with $s_i = |t_i|$. \textbf{Top figure (a)}: the red plane separates separable (left) and entangled (right) T-states. The sandwiched blue surface  corresponds to the necessary condition for EPR-steerability generated by  our deterministic   LHS model in Sec.~4B: all  T-states  to the left of this surface are not EPR-steerable. We conjecture that this condition is also sufficient, i.e., that all states to the right of the blue surface are EPR-steerable. For comparison, the green plane corresponds to the  sufficient condition for EPR-steerability in Eq.~(\ref{linsuff}) of section 5A: all T-states to the right of this surface are EPR-steerable.  Only a portion of the surfaces are shown, as they are symmetric under permutations of $s_1,s_2,s_3$. \textbf{Bottom figure (b)}: Cross section through the top figure at $s_1=s_2$, where the necessary condition can be determined analytically (see Sec.~4D). The additional black dashed curve corresponds to the non-linear sufficient condition for EPR-steerability in Eq. \eqref{nonlinsuff}.}
\label{All_Tstate_Bounds}
\end{center}
\end{figure}

\subsection{Special cases}

When $|t_1| = |t_2|$ we can solve Eq.~(\ref{surface}) explicitly, because the normalisation constant $N_T$ simplifies. The corresponding equation of the $s_3$ semiaxis, in terms of $u := s_3/s_1=s_3/s_2$, is given by
\begin{align} \label{deg}
\hspace{-0.2cm}s_3=
\left\{\begin{array}{cc}
\left[1 + \frac{\mathrm{arctan}(\sqrt{u^{-2} - 1})}{u^2\sqrt{u^{-2} - 1}} \right] ^{-1} & u < 1,  \\
\left[1 - \frac {\sqrt{1-u^{-2}}}{2(u^2-1)}\ln\frac{|1-\sqrt{1-u^{-2}}|}{1+\sqrt{1-u^{-2}}} \right] ^{-1} & u > 1,
\end{array}
\right.
\end{align}
and $s_3=\frac12$ for $u=1$.  Fig.~1(b) displays this analytic EPR-steerable curve through the T-state subspace $|t_1|=|t_2| \Leftrightarrow s_1 = s_2$, showing more clearly the different correlation regions.

The symmetric situation $s_1=s_2=s_3$ corresponds to Werner states.  Our deterministic LHS model is for $s_1=s_2=s_3=1/2$ in this case, which is known to represent the EPR-steerable boundary for Werner states \cite{jones07}. Thus, our model is certainly optimal for this class of states.

\section{Sufficient conditions for EPR-steerability}\label{sec_sufficient}

In the previous section a strong necessary condition for the EPR-steerability of T-states was obtained, corresponding to the boundary defined in Eq.~(\ref{surface}) and depicted in Fig.~1. While we have conjectured that this condition is also sufficient, it is not actually known if all T-states above this boundary are EPR-steerable. Here we give two sufficient general conditions for EPR-steerability, and apply them to T-states.

These conditions are examples of EPR-steering inequalities, i.e., statistical correlation inequalities that must be satisfied by any LHS model for Bob \cite{cavalcanti09}. Thus, violation of such an inequality immediately implies that Alice and Bob must share an EPR-steerable resource.

Our first condition is based on a new EPR-steering inequality for the spin covariance matrix, and the second on a known nonlinear EPR-steering inequality \cite{jones11}. Both EPR-steering inequalities are further of interest in that they are asymmetric under the interchange of Alice and Bob's roles.

\subsection{Linear asymmetric EPR-steering inequality}

Suppose Alice and Bob share a two-qubit state with spin covariance matrix $C$ given by
\begin{equation} \label{cjk}
C_{jk}:= \langle \sigma_j\otimes\sigma_k\rangle - \langle \sigma_j\otimes\I\rangle\,\langle \I\otimes\sigma_k\rangle = T_{jk}-a_jb_k,
\end{equation}
and  that each can measure any Hermitian observable on their qubit. We show in Appendix~\ref{app:cov} that, if there is an LHS model for Bob, then the singular values $c_1$, $c_2$, $c_3$ of the spin covariance matrix must satisfy the linear EPR-steering inequality
\begin{equation} \label{lin}
c_1 + c_2 + c_3 \leq \frac{3}{2}\sqrt{1 - b^2}.
\end{equation}

From $C=T-\ve a\ve b^\intercal$, and using $\ve a=\ve b=\ve 0$ and $s_j=|t_j|$ for T-states, it follows immediately that one has the simple {\it sufficient} condition
\begin{equation} \label{linsuff}
s_1 +s_2+s_3 > \frac{3}{2}
\end{equation}
for the EPR-steerability of T-states (by either Alice or Bob). The boundary of T-states satisfying this condition is plotted in Figs.~1 (a) and~(b), showing that the condition is relatively strong. In particular, it is a tangent plane to the necessary condition at the point
corresponding to Werner states (which we already knew to be a point on the true boundary of EPR-steerable states). However, in some parameter regions a stronger condition can be obtained, as per below.

\subsection{Nonlinear asymmetric EPR-steering inequality}

Suppose Alice and Bob share a two-qubit state as before, where Bob can measure the observables $\I\otimes\sigma_3$, $\I\otimes\sigma_\phi$ on his qubit, with $\sigma_\phi:=\sigma_1\cos \phi + \sigma_2\sin\phi$, for any $\phi\in[0,2\pi]$, and Alice can measure corresponding Hermitian observables $A_3\otimes\I, A_\phi\otimes\I$ on her qubit, with outcomes labelled by $\pm1$.  It may then be shown that any LHS model for Bob   must satisfy the EPR-steering inequality~\cite{jones11}
\begin{eqnarray*}\label{non-linear_inequality}
&&\frac{1}{\pi} \int_{-\pi/2}^{\pi/2} \langle A_\phi \otimes\sigma_\phi \rangle \, d\phi  \\
 &~&~~~~\leq \frac{2}{\pi} \left[p_{+} \sqrt{1 - \langle {\I\otimes\sigma}_{3} \rangle_{+}^{2}} + p_{-} \sqrt{1 - \langle {\I\otimes\sigma}_{3} \rangle_{-}^{2}}\right],
\end{eqnarray*}
where $p_{\pm}$ denotes the probability that Alice obtains result $A_3=\pm 1$, and $\langle {\I\otimes\sigma}_{3} \rangle_{\pm}$ is Bob's corresponding conditional expectation value for $\I\otimes\sigma_3$ for this result.

As per the first part of Sec.~4A, we may always choose a representation in which the spin correlation matrix $T$ is diagonal, i.e., $T={\rm diag}[t_1,t_2,t_3]$, without loss of generality. Making the choices $A_3=\sigma_3$ and $A_\phi=\sigma_1 ({\rm sign}\, t_1)\cos\phi+ \sigma_2 ({\rm sign}\, t_2)\sin\phi$ in this representation, then $p_\pm$ and $\langle {\I\otimes\sigma}_{3} \rangle_{\pm}$ are given by $p_{\ve e}$ and the third component of $\ve b(\ve e)$ in  Eqs.~(\ref{pe}) and (\ref{qsteerB}),  respectively, with $\ve e=(0,0,\pm1)^\intercal$. Hence, the above inequality  simplifies to
\begin{align} \label{nlin}
|t_1|+|t_2| \leq \frac{2}{\pi}&\left[ \sqrt{(1+a_3)^2-(t_3+b_3)^2}\right. \nonumber\\
&~~~~ \left.+ \sqrt{(1-a_3)^2-(t_3-b_3)^2}\,\right],
\end{align}
where $a_3$ and $b_3$ are the third components of Alice and Bob's Bloch vectors $\ve a$ and $\ve b$.

For T-states, recalling that $s_i\equiv |t_i|$, the above inequality simplifies further, to the nonlinear inequality
\[ f(s_1,s_2,s_3):= s_1+s_2 -\frac{4}{\pi}\sqrt{1-s_3^2} \leq 0. \]
Hence, since similar inequalities can be obtained by permuting $s_1, s_2,s_3$, we have the {\it sufficient} condition
\begin{equation} \label{nonlinsuff}
\max\{ f(s_1,s_2,s_3), f(s_2,s_3,s_1), f(s_3,s_1,s_2)\} > 0
\end{equation}
for the EPR-steerability of T-states.  The boundary of T-states satisfying this condition is plotted in Fig.~1(b) for the case $s_1=s_2$.  It is seen to be stronger than the linear condition in Eq.~(\ref{linsuff}) if one semiaxis is sufficiently large. The region below both sufficient conditions is never  far above the smooth curve of our necessary condition, supporting our conjecture that the latter is the true boundary.

\section{Recapitulation and future directions}

In this paper we have considered steering for the set of two-qubit states with maximally mixed marginals (`T-states'), where
Alice is allowed to make arbitrary projective measurements on her qubit. We have constructed a LHV--LHS model (LHV for Alice, LHS for Bob), which describes measurable quantum correlations for all separable, and a large portion of non-separable, T-states. That is, this model reproduces the steering scenario, by which Alice's measurement collapses
Bob's state to a corresponding point on the  surface of the quantum steering ellipsoid.
Our model is constructed using the steering ellipsoid, and coincides with the optimal LHV--LHS model
for the case of Werner states.  Furthermore, only a small  (and sometimes vanishing)
gap remains between  the set of T-states that are provably non-steerable by our LHV--LHS model,
and the set that are provably steerable by the two steering inequalities that we derive. As such, we conjecture that this LHV--LHS model is in fact optimal for T-states. Proving this, however, remains an open question.

A natural extension of this work is to consider LHV--LHS models for arbitrary two-qubit states. How can knowledge of their steering ellipsoids be incorporated into such LHV--LHS models? Investigations in this direction have already begun, but the situation is far more complex when Alice and Bob's Bloch vectors have nonzero magnitude and the phenomenon of ``one-way steering'' may arise \cite{1waysteer}.

Finally, our LHV--LHS models apply to the case where Alice is restricted to measurements of Hermitian observables. It would be of great interest to generalize these to arbitrary POVM measurements. However, we note that this is a very difficult problem even for the case of two-qubit Werner states \cite{werner14}. Nevertheless, the steering ellipsoid is a depiction of all collapsed states, including those arising from POVMs (they give the interior points of the ellipsoid) and perhaps this can provide some intuition for how to proceed with this generalisation.

\begin{acknowledgments}
SJ would like to thank David Jennings for his early contributions to this project.
SJ is funded by EPSRC grant EP/K022512/1. This work was supported by the Australian Research Council Centre of Excellence CE110001027 and the European Union Seventh Framework Programme (FP7/2007-2013) under grant agreement n$^{\circ}$ [316244].
\end{acknowledgments}

\appendix

\section{Details of the deterministic LHS model}

The family of T-states described by our deterministic LHS model in Sec.~4B corresponds to the surface defined by either of Eqs.~(\ref{surface}) and (\ref{surface2}).  This is a consequence of the following theorem, proved further below.
\begin{thm}
For any full-rank diagonal matrix $T$ and nonzero vector ${\ve v}$ one has
\begin{align}
\int_{\ve n\cdot\ve v\geq 0} \frac{\ve n\,\dd^2\ve n}{(\ve n^\intercal T^{-2} \ve n)^{2}} = \frac{\pi |\!\det T|\, T^2 \ve v}{|T\ve v|}.
\end{align}
\end{thm}

Note that substitution of Eq.~(\ref{re}) into constraint~(\ref{con}) immediately yields Eq.~(\ref{surface}) via the theorem (with $\ve v=T^{-1}\ve e$).
Further, taking the dot product of the integral in the theorem with $\ve v$, multiplying by $N_T$, and integrating  $\ve v$ over the unit sphere, yields (reversing the order of integration)
\begin{align*}
\int \dd^2\ve n\,P(\ve n)\int_{\ve n\cdot\ve v\geq 0}\dd^2\ve v\, \ve v\cdot\ve n=  \pi,
\end{align*}
whereas carrying out the same operations on the righthand side of the theorem yields $\pi N_T|\det T|\int \sqrt{\ve v^\intercal T^2\ve v}\dd^2\ve v$.  Equating these immediately implies the equivalence of Eqs.~(\ref{surface}) and (\ref{surface2}) as desired.  An explicit analytic formula for the normalisation constant $N_T$ is given at the end of this appendix.

\begin{proof}
First, define
 $Q = T^{-2} \in GL(3,\mathbb{R})$; that is,
 \begin{align}
 Q = \diag(a,b,c) = (t_1^{-2}, t_2^{-2}, t_3^{-2}),
 \end{align}
 and
\begin{align} \label{qv}
\ve q (\ve v) :=  \int_{\ve n\cdot\ve v \geq 0} \frac{\ve n \dd^2 \ve n}{(\ve n^\intercal Q \ve n)^{2}}.
\end{align}
Noting $\ve v$ in the theorem may be taken to be a unit vector without loss of generality, we
will parameterise the unit vectors $\ve n$ and $\ve v$  by
\begin{align}
\ve n &= (\sin \theta \cos \phi,\sin\theta\sin\phi,\cos\theta)^\intercal,\\
\ve v &= (\sin \alpha \cos \beta,\sin\alpha\sin\beta,\cos\alpha)^\intercal,
\end{align}
with $\theta, \alpha \in [0,\pi]$ and  $\phi, \beta \in [0,2\pi)$.  Thus, $\dd^2 \ve n \equiv \sin\theta\dd\theta\dd\phi$.
Further, without loss of generality, it will be assumed that $\ve{v}$ points into
the northern hemisphere, so that $\cos \alpha\geq 0$. Then $\alpha\in \lbrack 0,\pi /2]$ and $\beta\in \lbrack 0,2\pi )$.

The surface of integration is a hemisphere bounded by the great circle $%
\ve n\cdot \ve v=\,0$. In the simple case where $\ve{v}%
=(0,0,1)^{T}$, the boundary curve has the parametric form
$(x,y,z)=(\cos \gamma,\sin \gamma,0)$ for $\gamma\in ( 0,2\pi )$.
Hence, the boundary curve in the generic case can be constructed by applying the
orthogonal operator $R$, that rotates $\ve{v}$ from $(0,0,1)^{T}$ to $(\sin \alpha \cos \beta,\sin\alpha\sin\beta,\cos\alpha)^\intercal$, to the vector $(\cos \gamma,\sin \gamma,0)^{T}$.
That is,
\begin{align*}
R&=\left(
\begin{array}{ccc}
\cos \beta & -\sin \beta & 0 \\
\sin \beta & \cos \beta & 0 \\
0 & 0 & 1%
\end{array}%
\right) \left(
\begin{array}{ccc}
\cos \alpha & 0 & \sin \alpha \\
0 & 1 & 0 \\
-\sin \alpha & 0 & \cos \alpha%
\end{array}%
\right) \\
&=\left( \allowbreak
\begin{array}{ccc}
\cos \alpha\cos \beta & -\sin \beta & \sin \alpha\cos \beta \\
\cos \alpha\sin \beta & \cos \beta &\sin \alpha \sin \beta \\
-\sin \alpha& 0 & \cos \alpha%
\end{array}%
\right),
\end{align*}
and the boundary curve has the form
\begin{align*}
\left(
\begin{array}{c}
x \\
y \\
z%
\end{array}%
\right) \!=\!R\left(
\begin{array}{c}
\cos \gamma \\
\sin \gamma \\
0%
\end{array}%
\right)\! =\!\left( \allowbreak
\begin{array}{c}
\cos \alpha\cos \beta\cos \gamma-\sin \beta\sin \gamma \\
\cos \alpha\sin \beta\cos \gamma+\cos \beta\sin \gamma \\
-\sin \alpha\cos \gamma%
\end{array}%
\right).
\end{align*}

For the purposes of integrating over the hemisphere, it is convenient to
vary $\phi $ from $0$ to $2\pi $ and $\theta $ from $0$ to its value $\chi
(\phi )$ on the boundary curve. From the above expression for the boundary,
and using $z=\cos \theta $ and $y/x=\tan \phi $, it follows that
$\cos \chi =-\sin \alpha\cos \gamma$
and
$(\cos \alpha\sin \beta\cos \gamma+\cos \beta\sin \gamma)\cos \phi =(\cos \alpha\cos \beta\cos \gamma-\sin \beta\sin \gamma)\sin \phi$.
The last equation be rearranged to read
$\cos \alpha\sin (\phi -\beta)\cos \gamma=\cos (\phi -\beta)\sin \gamma$,
and after squaring both sides this equation solves to give
\[
\cos \gamma=\pm \frac{\cos (\phi -\beta)}{[\cos ^{2}(\phi -\beta)+\cos ^{2}\alpha\sin
^{2}(\phi -\beta)]^{1/2}}.
\]%
Now, $\chi $ assumes its maximum value when $\phi =\beta$, which according to
the relation $\cos \chi =-\sin \alpha\cos \gamma$ and the fact that $\alpha \in [0,\pi/2]$ should correspond to $\gamma=0$. So we
take the upper sign in the last equation, yielding
\begin{align}
\label{coschi} \nonumber
\cos \chi &=\frac{-\sin \alpha\cos (\phi -\beta)}{[\cos ^{2}(\phi -\beta)+\cos ^{2}\alpha\sin
^{2}(\phi -\beta)]^{1/2}}\\
&= \frac{-\sin \alpha\cos (\phi -\beta)}{[\cos ^{2}\alpha+\sin
^{2}\alpha\cos ^{2}(\phi -\beta)]^{1/2}} .
\end{align}
It follows immediately that
\begin{align}
\label{sinchi}
\sin \chi =\frac{\cos \alpha}{[\cos ^{2}\alpha+\sin ^{2}\alpha\cos ^{2}(\phi -\beta)]^{1/2}},
\end{align}
with the choice of sign fixed by the fact that $\sin \chi \geq 0$ and (by
assumption) $\cos \alpha\geq 0$.

The surface integral for $\ve q(\ve v)$ in Eq.~(\ref{qv}) can now be written in the form:
\begin{equation} \label{qvnew}
\int_{0}^{2\pi }\int_{0}^{\chi (\phi )}\frac{(\sin \theta \cos \phi ,\sin
\theta \sin \phi ,\cos \theta )^{T}\sin \theta \,d\theta
\,d\phi}{(a\sin ^{2}\theta \cos ^{2}\phi +b\sin
^{2}\theta \sin ^{2}\phi +c\cos ^{2}\theta )^{2}} .
\end{equation}

To evaluate the the third component of $\ve q(\ve v)$, note that the integral over $\theta $,
\[
\int_{0}^{\chi (\phi )}\frac{\sin \theta \cos \theta \dd \theta }{(a\sin ^{2}\theta
\cos ^{2}\phi +b\sin ^{2}\theta \sin ^{2}\phi +c\cos ^{2}\theta )^{2}},
\]%
can be evaluated explicitly by making the substitution $w=\sin ^{2}\theta $,
as $\int (A +B w)^{-2}dw=-B ^{-1}(A +B w)^{-1}$ for
any $B \neq 0$, yielding
\begin{align*}
\frac{1}{2c}\frac{\sin ^{2}\chi }{a\sin ^{2}\chi \cos ^{2}\phi +b\sin
^{2}\chi \sin ^{2}\phi +c\cos ^{2}\chi }.
\end{align*}
After inserting the expressions for $\cos \chi $ and $\sin \chi $ derived
earlier, we have
\begin{eqnarray*}
&&\int_{0}^{\chi (\phi )}\frac{\sin \theta \cos \theta }{(a\sin ^{2}\theta
\cos ^{2}\phi +b\sin ^{2}\theta \sin ^{2}\phi +c\cos ^{2}\theta )^{2}}%
\,d\theta  \\
&=&\frac{1}{2c}\frac{\cos ^{2}\alpha}{a\cos ^{2}\alpha\cos ^{2}\phi +b\cos ^{2}\alpha\sin
^{2}\phi +c\sin ^{2}\alpha\cos ^{2}(\phi -\beta)}.
\end{eqnarray*}
We now need to integrate the last expression over $\phi $. Introducing new constants
\begin{align*}
l &=a\cos ^{2}\alpha+c\sin ^{2}\alpha\cos ^{2}\beta,\\
m &=b\cos^{2}\alpha+c\sin ^{2}\alpha\sin ^{2}\beta,\\
n &=c\sin ^{2}\alpha\sin \beta\cos \beta,
\end{align*}
the full surface integral simplifies to a form that may be evaluated by Mathematica (or by contour integration over the unit circle in the complex plane):
\begin{eqnarray} \nonumber
\label{int1}
&&\int_{0}^{2\pi }\int_{0}^{\chi (\phi )}\frac{\sin \theta \cos \theta d\theta \,d\phi }{%
(a\sin ^{2}\theta \cos ^{2}\phi +b\sin ^{2}\theta \sin ^{2}\phi +c\cos
^{2}\theta )^{2}} \\ \nonumber
&=&\frac{\cos ^{2}\alpha}{2c}\int_{0}^{2\pi }\frac{d\phi }{l \cos ^{2}\phi +m
\sin ^{2}\phi +2n \sin \phi \cos \phi }\\ \nonumber
&=& \pm\frac{\cos ^{2}\alpha}{2c} \frac{2\pi}{\sqrt{lm-n^2}}.
\end{eqnarray}%
The indeterminate sign here  is fixed by examining the case $\alpha=0$ and $%
a=b=c$, for which $\chi (\phi )=\pi /2$ and the integrand reduces to $a^{-2}\sin
\theta \cos \theta $, which integrates to give $\pi a^{-2}$. So,
unsurprisingly, we choose the positive  sign. This  yields the third component of surface integral to be
\begin{align} \label{qv3}
[\ve q(\ve v)]_3 = \frac{\pi \cos \alpha}{c[ab\cos ^{2}\alpha+c(a\sin ^{2}\beta+b\cos ^{2}\beta)\sin
^{2}\alpha]^{1/2}}.
\end{align}

The integrals over $\theta $ in the remaining two components of $\ve q(\ve v)$  in Eq.~(\ref{qvnew}) are unfortunately not so straightforward.
However, there is a simple trick that allows us to calculate both surface
integrals explicitly, and that is to differentiate the integrals with
respect to the parameters $\alpha$ and $\beta$. Since the only dependence on $\alpha$ and $%
\beta$ comes through the function $\chi (\phi)$, this eliminates the need to integrate
over $\theta $. In fact we only need to differentiate with respect to one of these parameters, choose $\alpha$. To see this, note that
\begin{align*}
&\frac{\partial }{\partial \alpha}\int_{0}^{2\pi }\int_{0}^{\chi (\phi )}\frac{%
(\cos \phi ,\sin \phi )\sin ^{2}\theta \,d\theta \,d\phi}{(a\sin ^{2}\theta \cos ^{2}\phi
+b\sin ^{2}\theta \sin ^{2}\phi +c\cos ^{2}\theta )^{2}}  \\
&=\int_{0}^{2\pi }\frac{(\cos \phi ,\sin \phi )\sin ^{2}\chi }{(a\sin
^{2}\chi \cos ^{2}\phi +b\sin ^{2}\chi \sin ^{2}\phi +c\cos ^{2}\chi )^{2}}\,%
\frac{\partial \chi }{\partial \alpha}d\phi,
\end{align*}%
where $\partial \chi /\partial \alpha$ can be evaluated by making use of the equations \eqref{coschi} and \eqref{sinchi}.

In fact,
\begin{align*}
-\sin \chi \,\frac{\partial \chi }{\partial \alpha}&=\,\frac{\partial }{\partial \alpha}%
\left(\frac{-\sin \alpha\cos (\phi -\beta)}{[\cos ^{2}\alpha+\sin ^{2}\alpha\cos ^{2}(\phi -\beta)]^{1/2}%
}\right)
\end{align*}
\begin{align*}
=\frac{-\cos \alpha\cos (\phi -\beta)}{[\cos ^{2}\alpha+\sin ^{2}\alpha\cos ^{2}(\phi
-\beta)]^{3/2}}.
\end{align*}
Inserting the last two equations and the expressions for $\sin\chi$ and $\cos\chi$ into the integrals above, and using the constants $l, m$ and $n$ defined earlier, then gives:
\begin{widetext}
\begin{eqnarray} \nonumber
&&\frac{\partial }{\partial \alpha}\int_{0}^{2\pi }\int_{0}^{\chi (\phi )}\frac{%
(\cos \phi ,\sin \phi )\sin ^{2}\theta \dd\theta \,\dd\phi}{(a\sin ^{2}\theta \cos ^{2}\phi
+b\sin ^{2}\theta \sin ^{2}\phi +c\cos ^{2}\theta )^{2}}   \\
~&=&\cos ^{2}\alpha\int_{0}^{2\pi }\frac{(\cos \phi ,\sin \phi )\cos (\phi -\beta)}{%
[a\cos ^{2}\phi \cos ^{2}\alpha+b\sin ^{2}\phi \cos ^{2}\alpha+c\sin ^{2}\alpha\cos
^{2}(\phi -\beta)]^{2}}\dd\phi  \nonumber \\
\label{xy1}
~&=&\cos ^{2}\alpha\int_{0}^{2\pi }\frac{(\sin \beta\sin \phi \cos \phi +\cos \beta\cos
^{2}\phi ,\sin \beta\sin ^{2}\phi +\cos \beta\sin \phi \cos \phi )}{(l \cos
^{2}\phi +m \sin ^{2}\phi +2n \sin \phi \cos \phi )^{2}}\dd\phi .
\end{eqnarray}%
\end{widetext}

Consequently, there are three separate integrals we need to evaluate and these can be done in Mathematica (or by complex contour integration):
\begin{eqnarray*}
\label{int2}
\int_{0}^{2\pi }  \frac{(\sin^2\phi,\cos ^{2}\phi,\sin\phi\cos\phi) \, d\phi }{(l\cos ^{2}\phi +m \sin ^{2}\phi +2n \sin \phi \cos \phi )^{2}}
=  \frac{\pi(l,m,-n)}{(lm-n^2)^{3/2}} .
\end{eqnarray*}
Using the values we have for $l,m,n$ we substitute these back into equation \eqref{xy1} and integrate over $\alpha$ to obtain
\begin{align} \nonumber
\vspace{-1cm}
&[\ve q(\ve v)]_1 =\pi\int \frac{\cos ^{2}\alpha(m\cos \beta-n\sin \beta)  }{(lm-n^2)^{3/2}}d\alpha
\end{align}
\begin{align}
\label{qv1}
&=\frac{a^{-1}\pi \sin\alpha\cos\beta}{[ab \cos^2\alpha + c\sin^2\alpha(b\cos^2\beta + a\sin^2\beta)]^{1/2}},
\end{align}
and
\begin{align} \nonumber
&[\ve q(\ve v)]_2 =\pi\int \frac{\cos ^{2}\alpha(l\sin \beta -n\cos \beta) }{(lm-n^2)^{3/2}}d\alpha\\ \label{qv2}
&=\frac{b^{-1}\pi \sin\alpha\sin\beta}{[ab \cos^2\alpha + c\sin^2\alpha(b\cos^2\beta + a\sin^2\beta)]^{1/2}}.
\end{align}
 The absence
of  integration constants can be confirmed by noting that these expressions
 vanish for $\alpha=0$ -- i.e., when the
vector $\mathbf{\ve v}$ is aligned with the $z$-axis -- as they should by
symmetry.
Note the denominators of Eqs. \eqref{qv1} and \eqref{qv2} simplify to
$abc (\ve v^\intercal Q^{-1} \ve v )$.
Combining this with Eqs.~(\ref{qv3}) and (\ref{qv1})-(\ref{qv2}), we have
\begin{align}
\ve q (\ve v) = \frac{ \pi Q^{-1} \ve v}{\sqrt{abc (\ve v^\intercal Q^{-1} \ve v)}},
\end{align}
and so setting $Q = T^{-2}$, the theorem follows as desired.
\end{proof}

Finally, the normalisation constant $N_T$  in Eq.~(\ref{surface}) may be analytically evaluated using Mathematica. Under the assumption that $|t_3| > |t_2| > |t_1|$, denote $a=|t_1|, b=|t_2|, c=|t_3|$.  We find
\begin{widetext}
\begin{align} \nonumber
N_T^{-1} &= \int_{\ve n\cdot\ve n=1} (\ve n^\intercal T^{-2} \ve n)^{-2} \dd^2\ve n
= \frac{2\pi}{abc(a+b)(b+c)(c^2-a^2)}  \\
&\times  \big(X+Y\left\{b(c-a)E[C]+a(b+c)K[C]+ib(c-a)(E[A_1,B]-E[A_2,B])+ic(a+b)(F[A_1,B]-F[A_2,B])\right\}\big),
\end{align}
\end{widetext}
where $F[\cdot, \cdot], E[\cdot,\cdot]$ are the elliptic integrals the first and second kind, $E[\cdot]$ is the complete elliptic integral and $K[\cdot]$ is the complete elliptic integral of the first kind, and
\[
A_1 = i \mathrm{arccsch}\left(\frac{a}{\sqrt{c^2-a^2}} \right),~~~~
A_2 = i\ln \left(\frac{b+c}{\sqrt{c^2-b^2}} \right),
\]
\[
B = \frac{a^2(c^2-b^2)}{b^2(c^2-a^2)} ,~~~~~~~~
C = \frac{c^2(b^2-a^2)}{b^2(c^2-a^2)},
\]
\[
X =c(c-a)[(a+c)(b+c)+ab],~~~~
Y=(a+b+c)\sqrt{c^2-a^2}.
\]
Thus, the normalisation constant $N_T$ has a rather non-trivial form. It is highly unlikely that we can invert it to express the EPR-steerability condition $2\pi N_T|\det T| =1$ as $c = g(a,b)$ where $g$ is some function of $a,b$,  other than in the special cases considered in Sec.~4D. In general, we must leave it as an implicit equation in $a,b,c$ (that is, of the $t_j$s).

\section{\label{app:cov}EPR-Steering inequality for spin covariance matrix}

To demonstrate the linear EPR-steering inequality in Eq.~(\ref{lin}), let $A_{\ve v}$ denote some dichotomic observable that Alice can measure on her qubit,  with outcomes labelled by $\pm1$, where $\ve v$ is any unit vector.  We will make a specific choice of $A_{\ve v}$ below.  Define the corresponding covariance function
\begin{equation} \label{cv}
C(\ve v):=\langle A_{\ve v} \otimes \ve v\cdot\ve\sigma\rangle -\langle A_{\ve v}\rangle\,\langle \ve v\cdot\ve\sigma\rangle .
\end{equation}
If there is an LHS model for Bob then, noting that one may take $p(a|x,\lambda)$in Eq.~(\ref{lhs}) to be deterministic without loss of generality, there are functions $\alpha_{\ve v}(\lambda)=\pm 1$ such that
$C(\ve v) =\sum_\lambda p(\lambda) [\alpha_{\ve v}(\lambda) - \bar{\alpha}_{\ve v}]\,[\ve n(\lambda)-\ve b]\cdot\ve v$,
where $\bar{\alpha}_{\ve v}=\sum_\lambda p(\lambda) \alpha_{\ve v}(\lambda)$, and the hidden state $\rho_B(\lambda)$ has corresponding Bloch vector $\ve n(\lambda)$.

Now, the Bloch sphere can be partitioned into two sets,
$S_+(\lambda)=\{\ve v : [\ve  n(\lambda)-\ve b]\cdot\ve v \geq 0\}$ and $S_-(\lambda)=\{\ve v : [ n(\lambda)-\ve b]\cdot\ve v <0\}$,
for each value of $\lambda$.  Hence, noting $-1-\bar{\alpha}_{\ve v}\leq \alpha_{\ve v}(\lambda) - \bar{\alpha}_{\ve v} \leq 1- \bar{\alpha}_{\ve v}$, $\int  C(\ve v)\,d^2\ve v$ is equal to
\begin{eqnarray*}
&~& \sum_\lambda p(\lambda)\left\{ \int_{S_+(\lambda)}\dd^2\ve v\, [\alpha_{\ve v}(\lambda) - \bar{\alpha}_{\ve v}]\,[\ve n(\lambda)-\ve b]\cdot\ve v \right.\\
&~& \left. +  \int_{S_-(\lambda)} \dd^2\ve v\, [\alpha_{\ve v}(\lambda) - \bar{\alpha}_{\ve v}]\,[\ve n(\lambda)-\ve b]\cdot\ve v \right\}
\end{eqnarray*}
\begin{eqnarray*}
 &\leq& \sum_\lambda p(\lambda)\left\{ \int_{S_+(\lambda)} \dd^2\ve v\, [1 - \bar{\alpha}_{\ve v}]\,[\ve n(\lambda)-\ve b]\cdot\ve v \right.\\
&~& -\left. \int_{S_-(\lambda)} \dd^2\ve v\, [1+ \bar{\alpha}_{\ve v}]\,[\ve n(\lambda)-\ve b]\cdot\ve v \right\}\\
&=& \sum_\lambda p(\lambda) \int \dd^2\ve v\, |[\ve n(\lambda)-\ve b]\cdot\ve v|
\end{eqnarray*}
\begin{eqnarray*}
&~& - \sum_\lambda p(\lambda) \int \dd^2\ve v\, \bar{\alpha}_{\ve v}\,[\ve n(\lambda)-\ve b]\cdot\ve v\\
&=&  \sum_\lambda p(\lambda) |\ve n(\lambda)-\ve b|\,\int \dd^2\ve v\, |\ve v\cdot \ve w(\lambda)|,
\end{eqnarray*}
where $\ve w(\lambda)$ denotes the unit vector in the $\ve n(\lambda)-\ve b$ direction, and the last line follows by interchanging the summation and integration in the second term of the previous line.

The integral in the last line can be evaluated for each value of $\lambda$ by rotating the coordinates such that $w(\lambda)$ is aligned with the $z$-axis, yielding
$\int \dd^2\ve v\, |\ve v\cdot \ve w(\lambda)| = \int\dd^2\ve v\,|v_3|  = \int_0^{2\pi} \dd\phi \int_0^\pi \dd\theta \sin \theta\, |\cos\theta| =2\pi$.
Hence, the above inequality can be rewritten as
\begin{eqnarray} \nonumber
\frac{1}{4\pi} \int \dd^2\ve v\, C(\ve v) &\leq& \frac{1}{2} \sum_\lambda p(\lambda)\,|\ve n(\lambda) -\ve b| \\ \nonumber
&\leq& \frac{1}{2}\left[\sum_\lambda p(\lambda)\,|\ve n(\lambda) -\ve b|^2\right]^{1/2} \\ \label{cav}
&\leq& \frac{1}{2}\sqrt{1-\ve b\cdot\ve b},
\end{eqnarray}
where the second and third lines follow using the Schwarz inequality and $|\ve n(\lambda)|\leq 1$, respectively.  Note, by the way, that the first inequality is tight for the case $\alpha_{\ve v}(\lambda)={\rm sign}\left( [\ve n(\lambda)-\ve b]\cdot\ve v\right)$.

Now, making the choice $A_{\ve v}=\ve u\cdot\ve \sigma$ with $u_j:={\rm sign}(C_{jj}) v_j$, one has from Eqs.~(\ref{cjk}) and (\ref{cv}) that
\begin{eqnarray*}
\int \dd^2\ve v\, C(\ve v) &=&\sum_{j,k} C_{jk}\,{\rm sign}(C_{jj}) \int \dd^2\ve v\,v_j\,v_k \\
&=& \sum_{j,k} C_{jk}\,{\rm sign}(C_{jj}) \frac{4\pi}{3}\delta_{jk} = \frac{4\pi}{3}\sum_j |C_{jj}| .
\end{eqnarray*}
Combining with Eq.~(\ref{cav}) immediately yields the EPR-steering inequality
\begin{equation}
\sum_j |C_{jj}| \leq \frac{3}{2} \sqrt{1-\ve b\cdot\ve b}.
\end{equation}
Finally, this inequality may similarly be derived in a representation in which local rotations put the
spin covariance matrix $C$ in diagonal form, with coefficients given up to a sign by the singular values of $C$ (similarly to the spin correlation matrix $T$ in Sec.~4A). Since $\ve b\cdot\ve b=b^2$ is invariant under such rotations, Eq.~(\ref{lin}) follows.

\end{document}